\documentclass[conference]{IEEEtran}
\IEEEoverridecommandlockouts
\usepackage{cite}
\usepackage{amsmath,amssymb,amsfonts}
\usepackage{algorithm}
\usepackage{algorithmic}
\usepackage{graphicx}
\usepackage{textcomp}
\usepackage{xcolor}
\def\BibTeX{{\rm B\kern-.05em{\sc i\kern-.025em b}\kern-.08em
    T\kern-.1667em\lower.7ex\hbox{E}\kern-.125emX}}
\DeclareMathOperator*{\minimize}{minimize}
\makeatletter
\newcommand{\linebreakand}{%
      \end{@IEEEauthorhalign}
      \hfill\mbox{}\par
      \mbox{}\hfill\begin{@IEEEauthorhalign}
    }
\makeatother
  
\begin{document}

\title{Clustering of the Blendshape Facial Model
\thanks{This work has received funding from the European Union's Horizon 2020 research and innovation program under the Marie Sklodowska-Curie grant agreement No 812912, and from strategic project NOVA LINCS (FCT UIDB/04516/2020). The work has also been supported in part by the Ministry of Education, Science and Technological Development of the Republic of Serbia (Grant No. 451-03-9/2021-14/200125)}}

\author{
\IEEEauthorblockN{Stevo Racković}
\IEEEauthorblockA{
\textit{Institute for Systems and Robotics} \\
\textit{Instituto Superior Técnico}\\
Lisbon, Portugal \\
stevo.rackovic@tecnico.ulisboa.pt
}
\and
\IEEEauthorblockN{Cláudia Soares}
\IEEEauthorblockA{
\textit{Computer Science Department} \\
\textit{NOVA School of Science and Technology}\\
Lisbon, Portugal \\
claudia.soares@fct.unl.pt}
\and 
\IEEEauthorblockN{Dušan Jakovetić}
\IEEEauthorblockA{
\textit{Department of Mathematics} \\
\textit{Faculty of Sciences UNS}\\
Novi Sad, Serbia \\
dusan.jakovetic@dmi.uns.ac.rs
}
\linebreakand
\IEEEauthorblockN{Zoranka Desnica}
\IEEEauthorblockA{
\textit{3Lateral, Epic Games Company}\\
zoranka.desnica@3lateral.com
}
\and
\IEEEauthorblockN{Relja Ljubobratović}
\IEEEauthorblockA{
\textit{3Lateral, Epic Games Company}\\
relja.ljubobratovic@3lateral.com 
}
}

\maketitle

\begin{abstract}
Digital human animation relies on high-quality 3D models of the human face---rigs. A face rig must be accurate and, at the same time, fast to compute. One of the most common rigging models is the blendshape model.
We present a novel approach for learning the inverse rig parameters at increased accuracy and decreased computational cost at the same time. It is based on a two fold clustering of the blendshape face model. Our method focuses exclusively on the underlying space of deformation and produces clusters in both the mesh space and the controller space---something that was not investigated in previous literature. This segmentation finds intuitive and meaningful connections between groups of vertices on the face and deformation controls, and further these segments can be observed independently. A separate model for solving the inverse rig problem is then learnt for each segment. Our method is completely unsupervised and highly parallelizable.
\end{abstract}



\section{Introduction}
\label{sec:intro}

 Animation of a human face is a challenging problem in the industry due to our sensitivity to changes of expressions and the uncanny valley effect \cite{6213238}. In order to achieve realistic facial deformations different approaches are considered, but the most widely adopted one is the blendshape model. 
Blendshape modeling is a well-established research topic \cite{lewis2014practice, choe2001analysis, li2010example, lewis2010direct}, and the related literature branches into several directions, covering creation of the blendshape basis \cite{li2010example, li2013realtime, lewis2014practice, neumann2013sparse}, solving inverse rig problem \cite{choe2001analysis, li2010example, joshi2006learning, yu2014regression, ccetinaslan2016position, holden2016learning, song2017sparse}, direct manipulation \cite{lewis2010direct} and segmentation of the face  \cite{choe2001analysis,joshi2006learning,song2017sparse,na2011local, tena2011interactive, fratarcangeli39fast,romeo2020data}.

\textbf{Contribution.}
 In this paper we introduce a novel approach to face segmentation procedure in the context of solving inverse rig problems. We propose a two-fold clustering of the face that is based exclusively on the blendshape model. In the first pass we cluster the vertices of the face mesh and in the second we form groups of controllers corresponding to each mesh cluster. To the best of our knowledge, this is the first paper that discusses a clustering of the deformation controllers.
 
 The whole procedure is unsupervised and demands only a blendshape matrix and two user specified parameters: the number of desired clusters $K$ and a proportion of tolerable overlapping between clusters $p$. The resulting clusters are suitable for learning inverse rig parameters in parallel.
 Our method produces intuitively correct face clusters and the experiments produced $15-35\%$ decrease in prediction error and at the same time over $30\%$ decrease in the number of used vertices when solving a rig based on the obtained clusters compared to a whole face approach.
 
\textbf{Literature Review.}
There is no original reference paper for the blendshape model, but a thorough introduction can be found in \cite{lewis2014practice, choe2001analysis, li2010example, lewis2010direct}. The blendshape basis is conventionally sculpted by hand. The works \cite{li2010example, li2013realtime} introduce an automatic framework that adapts a generic model to a face of a new character, while \cite{lewis2014practice, neumann2013sparse} propose extracting a basis from a dense set of face scans.

Next stage in the pipeline is adjusting activation weights to produce the animation---this is called inverse rig and represents the main bottleneck in production due to the time involved.  Automatic solutions for certain linear forms of the rig are proposed in \cite{choe2001analysis, li2010example, joshi2006learning, yu2014regression, ccetinaslan2016position}. In order to enhance the fidelity of expression, additional corrective blendshapes can be introduced, yielding a nonlinear problem---as studied in \cite{holden2016learning, song2017sparse}. One generalization of inverse rig learning is a problem of direct manipulation, that considers an interface allowing a user to drag vertices of the face directly in order to produce the desired expression \cite{lewis2010direct}.

 Another important direction of research is clustering of the face. It allows different regions of the face to get observed and processed independently or in parallel. Early works consider a simple split of the face into upper and lower sets of markers \cite{choe2001analysis}. Later, these models are sought to be automatic \cite{joshi2006learning,song2017sparse} or semi-automatic \cite{na2011local, tena2011interactive, fratarcangeli39fast}. All these works use either topological positions of vertices in the face or their correlation over animation sequence, completely neglecting the underlying blendshape model. To the best of our knowledge, the only approach to clustering based on the underlying deformation model is a recent work of \cite{romeo2020data}. That work aims for adding a secondary motion on top of blendshape animation, and for this reason the authors insist on a very small granularity of clusters, and also need to include information on the direction of deformation for each vertex. On the other hand, our main goal is solving the inverse rig (in parallel), hence we aim for relatively large clusters of vertices that are affected together. Additionally, we have a second fold of clustering, that is assigning a specific set of controllers to each mesh cluster.
 
\textbf{Paper Organization.}
 The rest of this paper is organized as follows. Section II introduces the notation and the main ideas of blendshape animation. Section III explains the proposed clustering method. The method is evaluated in Section IV and finally the paper is concluded with a discussion in Section V. 

\section{Blendshape Animation}
\label{sec:sub_intro}

Consider a sculpted face in the neutral position. It consists of $n$ vertices $\textbf{v}_1,...,\textbf{v}_n\in\mathbb{R}^{3}$ on the surface mesh. We unravel coordinates of each vertex and stack them into a single vector $\textbf{b}_0\in\mathbb{R}^{3n}$. Additionally, we have a basis of $m$ atomic deformations of the face (rising of a left brow, stretching of a right mouth corner, etc.) vectorized in the same manner to yield $\textbf{b}_1,...,\textbf{b}_m\in\mathbb{R}^{3n}$. We call each $\textbf{b}_i$ a blendshape or a controller. A blendshape matrix $\textbf{B}\in\mathbb{R}^{3n\times m }$ is then formed as a matrix whose columns are blendshape vectors $\textbf{B}:=[\textbf{b}_1,...,\textbf{b}_m]$. Now any feasible facial expression can be obtained as 
$$f_L(\textbf{c}) = \textbf{b}_0 + \textbf{B}\textbf{c}$$ where $\textbf{c}=[c_1,...,c_m]^T$ is a vector of activation weights for each blendshape\footnote{Complex animation models can have additional corrective terms that would yield a nonlinear rig function $f(\textbf{c})$ \cite{ song2017sparse}.}. Mapping $f_L:\mathbb{R}^m \rightarrow \mathbb{R}^{3n}$, from parameters $\textbf{c}$ into a mesh space, is called a rig.

Similarly, one can define an inverse rig problem, which is a common problem in animation. The inverse rig considers a reference mesh $\hat{\textbf{b}}\in\mathbb{R}^{3n}$ that is conventionally obtained as a 3D scan of an actor, and the task is to find an optimal configuration $\hat{\textbf{c}}$ so that $f_L(\hat{\textbf{c}})\approx \hat{\textbf{b}}$. Problem is often stated as a least squares 
$$\minimize_{\textbf{c}} \|f_L(\textbf{c})-\hat{\textbf{b}}\|_2^2$$
with possible constraints on the structure or sparsity of $\textbf{c}$.

%

\section{Clustering in Mesh and Controllers}
\label{sec:method}

Some local models propose fitting separate regions or marker points independently when solving the inverse rig, and then interpolate the values in between, but each marker point is fitted using a complete set of $m$ controllers \cite{na2011local}. The majority of the controllers in each face vertex are redundant (points of the left ear are not affected by the controllers for the right eye), so we would like instead to consider only relevant combinations of vertices and controllers. This can be achieved by clustering in both mesh and controller space. 

\begin{figure}
\centering
\centerline{\includegraphics[width=\linewidth]{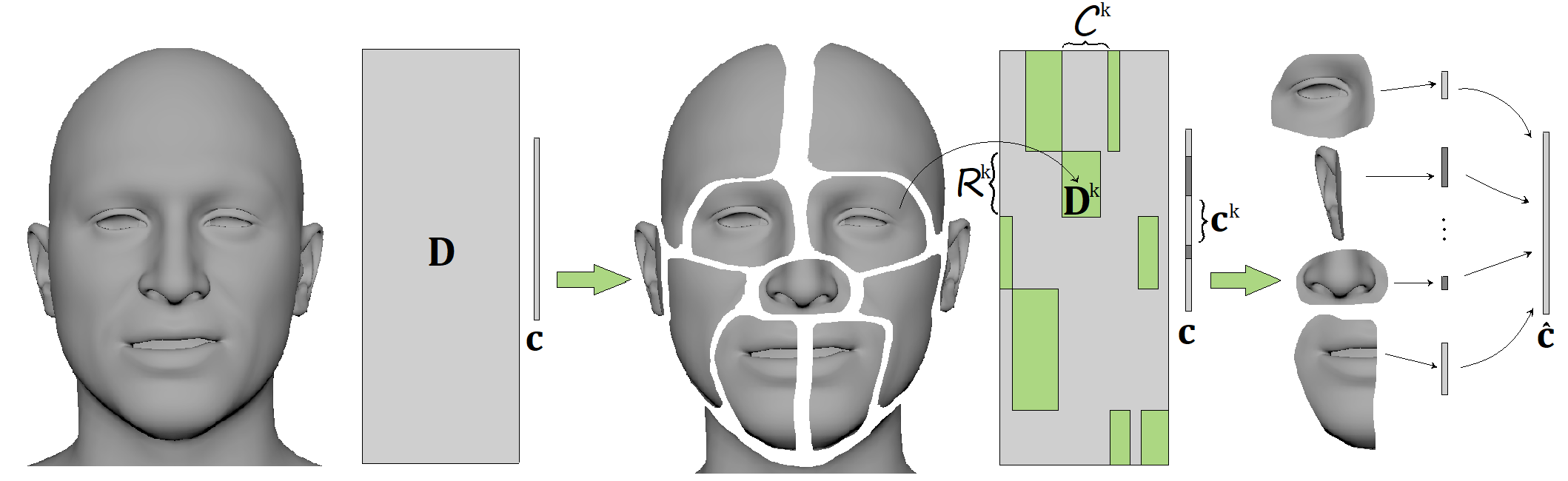}}
\caption{Scheme of the proposed approach. Originally we have a character face with a blendshape offset matrix \textbf{D}. We need to estimate a true value of weight vector \textbf{c}. Matrix \textbf{D} is clustered in both mesh (rows) and controller (columns) space, so the whole face model is divided into several submodels. Inverse rig problem is solved for each local cluster independently and the final results are aggregated into the prediction $\hat{\textbf{c}}.$  }
\label{fig:scheme} \vskip -.3cm
\end{figure}

More formally, we want to get a clustering $\{(\mathcal{R}^k,\mathcal{C}^k), k=1,...,K\}$ where $\mathcal{R}^k$ represents row clusters of a blendshape matrix $\textbf{B}$ i.e. $\mathcal{R}^k$ is a set of indices of mesh vertices that belong to the $k^{\text{th}}$ mesh cluster. $\mathcal{C}^k$ are the corresponding column clusters of matrix $\textbf{B}$, i.e. set of indices of controllers that are relevant for activating part of the mesh covered by $\mathcal{R}^k$. In this way we split a matrix $\textbf{B}$ into $K$ submatrices $\textbf{B}^k$ and hence we do not consider a single rig function $f_L(\cdot)$ but a separate rig function $f_L^k(\cdot)$ for each region of the head:
$$f_L^k(\textbf{c}^k) = \textbf{b}_0^k + \textbf{B}^k\textbf{c}^k.$$
Here, $\textbf{c}^k$ is a weight vector of controllers that belong to $\mathcal{C}^k$, $\textbf{b}_0^k$ are vertices of a neutral mesh that belong to $\mathcal{R}^k$ and $\textbf{B}^k$ is a submatrix of $\textbf{B}$ with rows in $\mathcal{R}^k$ and columns in $\mathcal{C}^k$ (See Fig. \ref{fig:scheme}).

We expect to find a natural structure that distinguishes parts of the face mesh and corresponding groups of controllers. Intuitively we would expect to have separate clusters for each eye and ear, for a mouth region, neck and ideally a cluster on the skull region containing inactive vertices that we could neglect in the process of inverse rig fitting. We will see later in results that the algorithm is able to discover this structure.

The method we propose consists of three steps that are explained in the following subsections. We first prepare a matrix of the offset magnitudes $\textbf{D}$ and normalize it (Section \ref{subsec:preprocessing}). Rows of the matrix are further clustered using K-means to produce mesh clusters, and we visit each controller to assign it to clusters where it shows a significant influence (Section \ref{subsection:clustering}). In the final phase, clusters that show high overlapping in controller space are merged together (Section \ref{subsection:overlapping}). The summarized procedure is given in Algorithm 1.

\subsection{Data Preparation}
\label{subsec:preprocessing}

As pointed out in \cite{neumann2013sparse}, if we work directly with the blendshape matrix $\textbf{B}$, we might end up with coordinates of a single vertex allocated into different clusters, and destroy the structure of the data. We want to guarantee that all the coordinates $v_l^x,v_l^y,v_l^z$ of a vertex $\textbf{v}_l$ in \textbf{B} will remain in the same cluster.
In order to allow for simultaneous clustering of controllers we consider a matrix of offset values $\textbf{D}\in\mathbb{R}^{n\times m}$. Columns $\textbf{d}_i$ of this matrix are obtained as an offset for each controller $i$: $${d}_i^l = \big\|[{b}_i^{3l},{b}_i^{3l-1},{b}_i^{3l-2}]\big\|_2 \,,\,\text{   for } l=1,...,n.$$
Here $b_i^{3l-2}$ represents entry of a blendshape $\textbf{b}_i$ that corresponds to $x$ coordinate of the vertex $\textbf{v}_l$. Similarly, superscripts $3l-1$ and $3l$ correspond to $y$ and $z$ coordinates of $\textbf{v}_l$. 

Some blendshapes produce larger offsets than others, so they can receive higher importance when clustering. To exclude this effect we additionally normalize matrix $\textbf{D}$. Each column is divided by its maximum, so that each blendshape has a maximum deformation offset value equal to 1. 

\subsection{Clustering}
\label{subsection:clustering}

\begin{figure}[]
\centering
\centerline{\includegraphics[width=.5\linewidth]{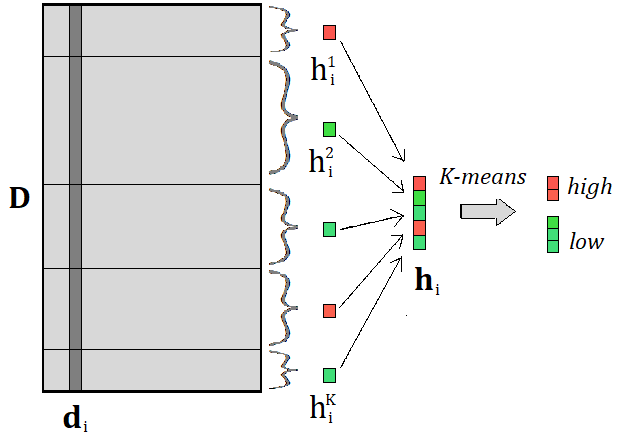}}
\caption{Assignment of controllers to clusters. Values of $\textbf{d}_i\in\mathbb{R}^n$ are averaged over mesh clusters to produce a vector $\textbf{h}_i\in\mathbb{R}^K$. Further K-means with $K=2$ is performed over $\textbf{h}_i$ to distinguish between high and low activation values. Controller $i$ is assigned to clusters that correspond to high activation.}
\label{fig:column_clust_diag} \vskip -.3cm
\end{figure}

The proposed clustering algorithm is two-fold. In the first step we consider clustering in the mesh space. For this we perform K-means\cite{bishop_2006_PRML} over rows of $\textbf{D}$ to obtain $K$ mesh clusters $\mathcal{R}^k$. In the second step we visit each controller $i$ to decide to which cluster it should be assigned. The idea is following: we take a column $\textbf{d}_i\in\mathbb{R}^n$ and compress it into $\textbf{h}_i\in\mathbb{R}^K$, with entries $$ {h}_i^k = \frac{\sum_{l \in \mathcal{R}^k}{d}_i^l}{|\mathcal{R}^k|}\,\,\text{ for }\,k=1,..,K.$$
Element $k$ of vector $\textbf{h}_i$ represents the average magnitude of the activation produced by the controller $i$ in a mesh cluster $\mathcal{R}^k$. We want to assign the controller only to those mesh clusters where the high activation is exhibited. (In general, most of the entries of $\textbf{h}_i$ will be close to zero, and only few of them will be considerably higher. These higher values or \textit{peaks} are the clusters we are interested in.) Hence, we perform one-dimensional K-means clustering with $K=2$ over the vector $\textbf{h}_i$. This will yield one segment with low activation values and the other with high values, so we assign controller $i$ to clusters that correspond to indices of high values of $\textbf{h}_i$ (See Fig. \ref{fig:column_clust_diag}). 

For each mesh cluster $\mathcal{R}^k$ we will get a controller cluster $\mathcal{C}^k$, which is a set of controller indices assigned to that mesh cluster. Final result is hence a set of mesh/controller cluster pairs $\{(\mathcal{R}^k,\mathcal{C}^k), \,\, k=1,...,K\}$.

\subsection{Overlapping Clusters}

\label{subsection:overlapping}

The proposed method allows for a single controller to be assigned to multiple clusters. This will in general produce overlapping pairs of controller clusters $\mathcal{C}^k$ and $\mathcal{C}^j$ (for some $j,k\in\{1,...,K\}$). This behavior is in accordance with the nature of face, but sometimes the overlapping proportion is very high, up to the point that cluster $\mathcal{C}^k$ is completely contained in cluster $\mathcal{C}^j$ or vice versa. Complex regions like mouth are especially susceptible to this, specifically if $K$ takes higher values. 

To address this, we introduce a final adjustment step. Consider a user specified parameter $0< p\leq1$ that represents a percentage of the overlap that we will allow. Now consider that we have an overlapping pair of controller clusters $\mathcal{C}^k$ and $\mathcal{C}^j$. In case that the intersection of two is high enough, i.e. 
$|\mathcal{C}^k \cap \mathcal{C}^j| > p\,\text{min}\{|\mathcal{C}^k|, |\mathcal{C}^j|\}$
we will merge two clusters into $(\mathcal{R}^k \cup \mathcal{R}^j,\mathcal{C}^k \cup \mathcal{C}^j).$

A complete pipeline is summarized in Algorithm 1 and Fig. \ref{fig:scheme} schematically describes the method. Notice that we do not consider a novel inverse rig solver---the novelty of our work is a two-fold segmentation of the blendshape face model that would enable using standard inverse rig solver in a distributed and localized manner. This leads to simultaneous improvements in both prediction accuracy and computational efficiency.

 \begin{algorithm}[H]
 \caption{Two-fold Clustering of the Face}
 \begin{algorithmic}[1]
 \renewcommand{\algorithmicrequire}{\textbf{Input:}}
 \renewcommand{\algorithmicensure}{\textbf{Output:}}
 \REQUIRE $\textbf{b}_1,...,\textbf{b}_m\in\mathbb{R}^{3n}$, $K\in\{1,...,n\}$, $p\in(0,1]$.
 \ENSURE  mesh/controller cluster pairs\\ $\{(\mathcal{R}^k,\mathcal{C}^k), k=1,...,K\}$.
  \STATE Create an offset matrix $\textbf{D}\in\mathbb{R}^{n\times m}$ with elements $${d}_i^l = \big\|[{b}_i^{3l},{b}_i^{3l-1},{b}_i^{3l-2}]\big\|_2,\,\,\,i=1,...,m,\,l=1,...,n.$$
  \STATE Perform K-means over rows of $\textbf{D}$ to obtain mesh clusters $\mathcal{R}^k$ for $k=1,...,K$.
  \FOR {$i = 1,...,m$}
    \STATE Compress $\textbf{d}_i\in\mathbb{R}^n$ into $\textbf{h}_i\in\mathbb{R}^K$ where $$ {h}_i^k = \frac{ \sum_{l \in \mathcal{R}^k}{d}_i^l}{|\mathcal{R}^k|}.$$
  \STATE perform K-means with $K=2$ over $\textbf{h}_i$. Indices of $\textbf{h}_i$ that correspond to the cluster with higher clustroid represent mesh clusters relevant for the controller $i$. Assign the cluster $i$ to relevant mesh clusters.
 \ENDFOR
 \FOR {$k,j\in\{1,...,K\}$}
   \IF {($p\,|\mathcal{C}^k \cap \mathcal{C}^j| > \text{min}\{|\mathcal{C}^k|, |\mathcal{C}^j|\}$)}
  \STATE merge clusters $k$ and $j$ into $(\mathcal{R}^k \cup \mathcal{R}^j,\mathcal{C}^k \cup \mathcal{C}^j).$
  \ENDIF
  \ENDFOR
  \RETURN $\{(\mathcal{R}^k,\mathcal{C}^k), k=1,...,K\}$
 \end{algorithmic}
 \end{algorithm}

\section{Evaluation of the Method}

Once we obtain the clusters, we can learn the inverse rig parameters in parallel. In a case when training data is available, Gaussian process regression (GPR) is a simple and effective approach for this \cite{holden2016learning}. We use GPR with a dot product kernel, but instead of using a complete head as an input and outputting a full set of controllers, we train $K$ separate models. Each model corresponds to one of $K$ clusters and takes as an input (output) only the mesh (controller) cluster of interest. Models are trained independently and in the end all the outputs are aggregated into a single output vector (the values of the controllers that correspond to more than one cluster are averaged). 

Since a method like this was not proposed before (solving inverse rig based on the mesh-controller clusters) we will use a whole face model as a benchmark\footnote{I.e. input is a whole head mesh without clustering, and the output is a complete vector of controller weights.}. A whole face model corresponds to $K=1$, and we will experiment with values of $K$ ranging up to 100. $K$ is a user specified parameter and there will be certain trade-offs for different choices. 
In order to evaluate the results we consider several common metrics for this problem. 

\subsection{Evaluation Metrics}

1. \textit{Controller Prediction Error (CE) and Mesh Prediction Error (ME).} These are the most straight forward and common metrics for evaluating inverse rig solutions. We train a GPR model \cite{rasmussen2003gaussian} taking face meshes as an input and outputting estimated controller weights. Those predictions produce the corresponding meshes, hence we have both controller and mesh prediction error. The mesh generating function is denoted by $f:\mathbb{R}^m\rightarrow\mathbb{R}^{3n}.$ If $\textbf{c}$ is a ground truth controller vector and $\hat{\textbf{c}}$ the estimated one, we have $$CE(\hat{\textbf{c}};\textbf{c}) = \frac{\|\hat{\textbf{c}}-\textbf{c}\|_2}{n},$$ $$ME(\hat{\textbf{c}};\textbf{c}) = \frac{\|f(\hat{\textbf{c}})-f(\textbf{c})\|_2}{n}.$$

2. \textit{Number of Considered Vertices (NCV).} Some choices of $K$ produce mesh clusters with no assigned controllers (eg. inactive vertices in the skull region), and we can neglect those when doing prediction. This reduces a computational cost for the task.

3. \textit{Maximal number of Controllers (CpC) and Vertices (VpC) per Cluster}. Number of vertices (controllers) is recorded for each nonempty cluster, and we extract the largest value. It gives us idea of the size for the largest subproblem.

\subsection{Data}

The experiments are performed over three human face datasets, provided by 3Lateral Studio for the purposes of this research. In \textit{Dataset 1} we have $m=147$ blendshapes, in \textit{Dataset 2},  $m=401$ and in \textit{Dataset 3}, $m = 154$. The number of vertices $n=5863$ is the same for the three sets. Each dataset is accompanied by a short animation sequence. These are subsampled and split into training and test sets. Training sets contain 231, 220 and 239 frames, and test sets contain 129, 180 and 61 frames for \textit{Dataset 1, Dataset 2} and \textit{Dataset 3} respectively. Each frame is represented by the ground truth activation weights and the corresponding mesh. GPR is trained over training frames and the presented results are obtained from test frames.

\subsection{Results} 

\begin{figure}[]
  \centering
  \centerline{\includegraphics[width=.7\linewidth]{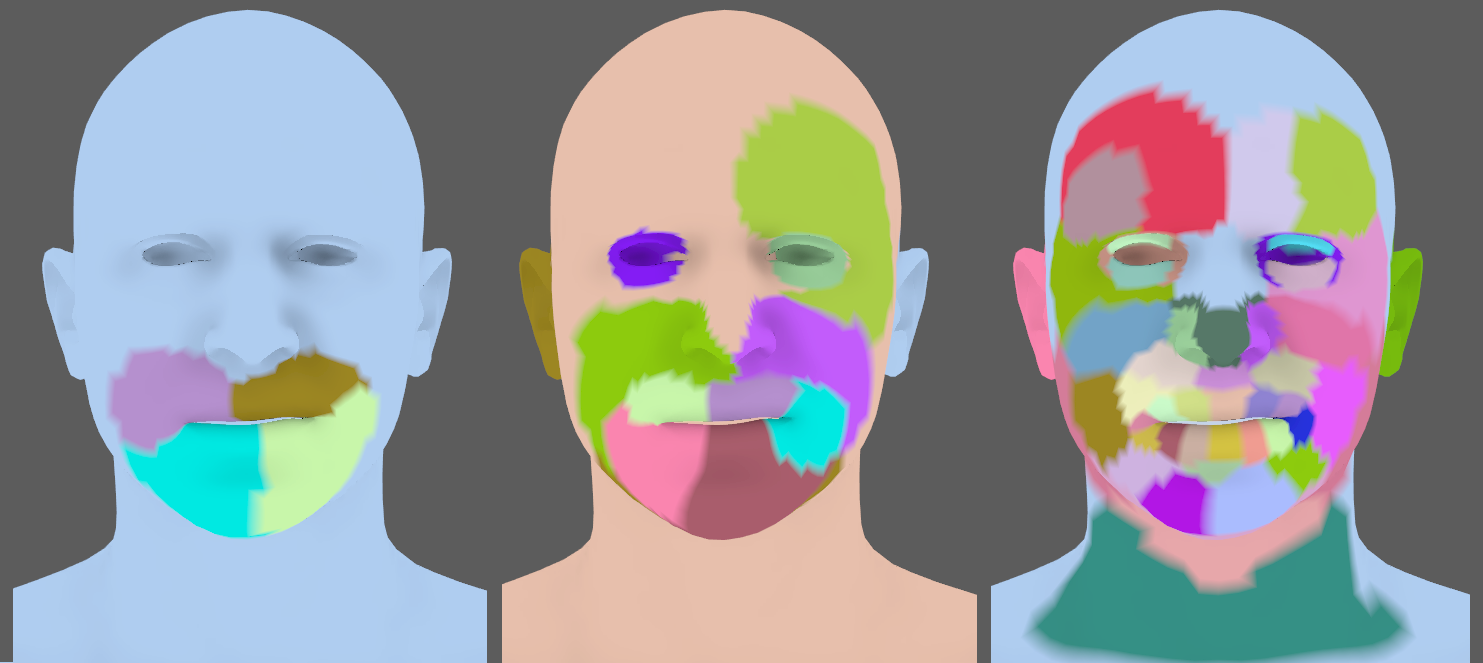}}
\caption{Mesh clusters for \textit{Dataset 1} with $K=5$, $K=13$ and $K=50$ respectively.}
\label{fig:first_meshes} \vskip -.5cm
\end{figure}

\begin{figure}[]
\centering
\centerline{\includegraphics[width=.45\linewidth]{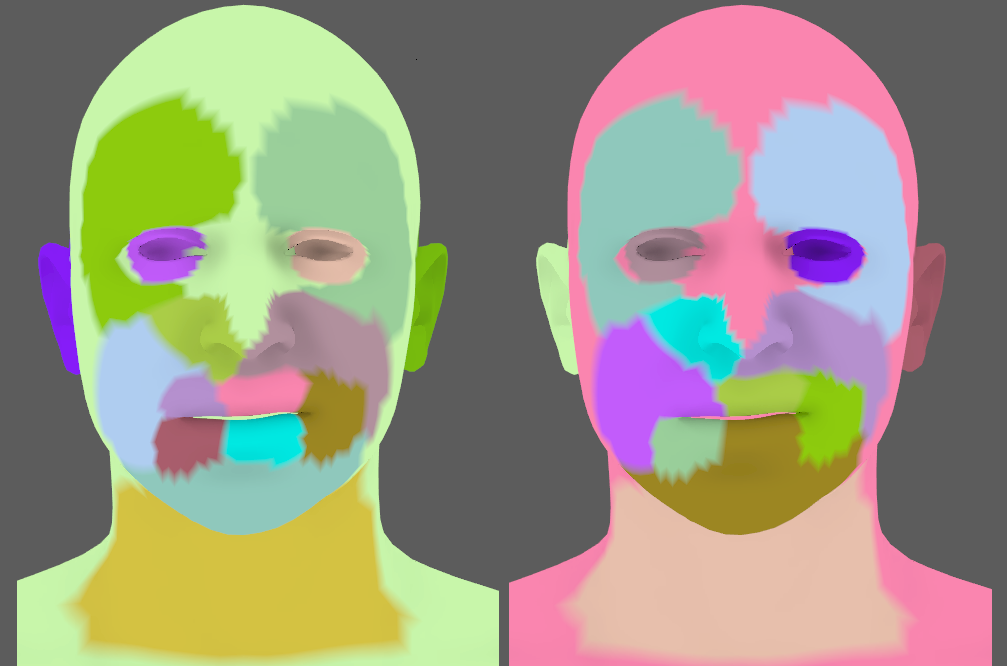}}
\caption{Mesh Clusters for \textit{Dataset 1}, $K=20$. Left: clusters before merging. Right: clusters after merging.}
\label{fig:second_meshes} \vskip -.3cm
\end{figure}

Let us first consider \textit{Dataset 1}. We try different choices of $K$ ranging from 2 to 100. In Fig. \ref{fig:first_meshes} we can see that the mouth region shows more detailed clusters, and that small values of $K$ will not yield relevant clusters for other parts of the face. For very large $K$ we get a large number of meaningless subdivisions. However, parameter $p$ serves as a form of regularizer---it will allow to merge most of the enforced small clusters. A crossvalidation indicates that a good choice of this parameter is $p=0.75$, hence we will use it in the rest of the experiments. Fig. \ref{fig:second_meshes} shows us the effects of merging, and we can see also the quantitative improvements in Table 1.

\begin{figure}[htb]
\centering
\centerline{\includegraphics[width=.9\linewidth]{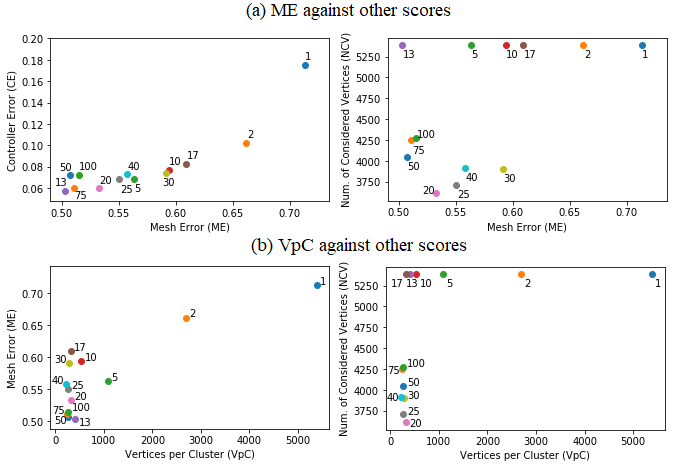}}
\caption{Evaluation scores for \textit{Dataset 1}. ME and CE are highly correlated, and tend to be higher for a low number of clusters ($K<5$). NCV drastically drops for $K\geq20$ while VpC exhibits exponential decay in $K$. Values of $K$ are represented as numbered colored dots.}
\label{fig:pairplot}
\end{figure}

Fig. \ref{fig:pairplot} shows values for pairs of evaluation metrics introduced above, for each $K$. We can see that CE and ME are highly correlated and that a whole face model exhibits a high error. Probably the most important is a relation between ME and NCV that gives us a trade-off between the size of the problem and the accuracy. It indicates that an optimal choice for this dataset would be $K=20$ or $K=50$ which is in accordance also with the relation between ME and VpC.

\begin{center}
\begin{tabular}{| c| c| c| c| c|}
\hline
               & ME    & CE    & CpC  & VpC   \\ 
   \hline
Before merging & 0.719 & 0.092 & 16.4 & 269.3 \\
\hline
After merging  & 0.532 & 0.059 & 15.9 & 316.8 \\
\hline
\end{tabular}\\
\vskip .1cm
\small{Table 1. Results for \textit{Dataset 1}, $K=20$ before and after merging.}\label{tablel}
\end{center}

\begin{figure}[]
\centering
\centerline{\includegraphics[width=.9\linewidth]{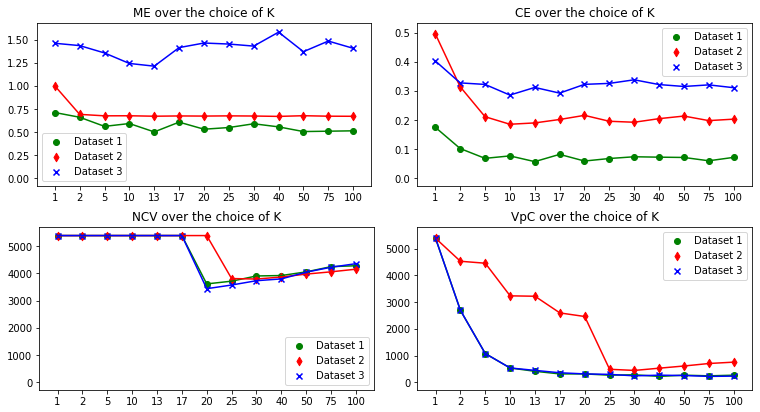}}
\caption{Different evaluation measures for three datasets. Modest choices of $K$ produce lowest ME and CE in each set, and NCV considerably drops for $K\approx20$.}
\label{fig:eval_models} \vskip -.5cm
\end{figure}

Evaluation results for all three datasets are presented in Fig. \ref{fig:eval_models}. Any choice of $K>1$ yields a lower CE than a whole face model, and we also have a significant decrease of the ME ($15-35\%$). Around the value of $K=20$ each dataset has a huge drop in NCV ($20-35\%$), while VpC is already very low. Hence, choosing $K\approx 20$ will yield a more accurate inverse rig solution at a lower computational cost, compared to a standard whole face approach. 

Finally we can conclude that solving inverse rig over a clustered face yields significant improvements over a whole face model with respect to each introduced metric of success. Exact choice of number of clusters $K$ still might depend on what user aims to minimize. If the accuracy is the only target, decision should be made based on ME (CE) values. If the size of a problem is a bottleneck, the best choice is the one where NCV drops. However, if a user has access to several processors, a parallel model can be implemented, and then the VpC and CpC are bottleneck, and slightly higher values of $K$ would work just as good. In most cases user will be interested in some trade-off between size and accuracy, and then some of the pairplots in Fig. \ref{fig:pairplot} should be consulted to check for the best joint improvement over $K$. 

\section{Conclusion and Discussion}

In this paper we investigated a novel approach to a blendshape face segmentation in a light of solving inverse rig problem. The method consists of a two-fold clustering of a deformation space (a blendsahpe matrix) that splits a face mesh into natural clusters and connects each mesh segment with relevant set of deformation controllers. Obtained mesh/controller cluster pairs then define submodels of a whole face blendshape model. These submodels are suitable for learning the inverse rig in parallel, and while the size of the problem is reduced the accuracy is increased simultaneously. 

The method proposed in this paper is fully automatic, and demands only two parameters to be specified: tolerance of the cluster overlapping $0<p\leq 1$, and a number of clusters $K$, that would depend on a character's face and later application of clusters. The experiments indicate that results are especially suitable for solving an inverse rig, as both the prediction error and the size of a matrix are reduced, and also the submodels can be learnt in parallel. The clustering itself is performed exclusively over a blendshape matrix, hence there is no need for additional training data.

In future experiments we plan to chose a different algorithm for inverse rig estimation. Gaussian process regression (GPR) is successful in solving inverse problem in animation \cite{holden2016learning}, but it cannot exploit the scaling properties of our clustering results to the full extent. Hence we shall investigate alternative approaches that might scale more favorably.

\bibliographystyle{IEEEbib}
\bibliography{references}

\end{document}